# STUDY OF THE PROBLEM OF REDUCING GREENHOUSE GAS EMISSIONS IN AGRICULTURAL PRODUCTION CZECH REPUBLIC

**Yekimov Sergiy**

Department of Trade and Finance

Faculty of Economics and Management

Czech University of Life Sciences Prague

Kamycka 129

165 00 Praha – Suchdol, Czech Republic

https://orcid.org/0000-0001-6575-2623

## Abstract

Agricultural production is the main source of greenhouse gas emissions, and therefore it has a great influence on the dynamics of changes in global warming.

The article investigated the problems faced by Czech agricultural producers on the way to reduce greenhouse gas emissions. The author analyzed the dynamics of greenhouse gas emissions by various branches of agriculture for the period 2000-2015. The author proposed the coefficient τ -covariances to determine the interdependence of the given tabular macroeconomic values. This indicator allows you to analyze the interdependence of macroeconomic variables that do not have a normal distribution. In the context of the globalization of the economy and the need to combat global warming in each country, it makes sense to produce primarily agricultural products that provide maximum added value with maximum greenhouse gas emissions.

# Introduction

The fight against climate change is one of the most important problems facing human civilization. According to a number of researchers [1,2,3], the delay in solving this problem will have catastrophic consequences for life on Earth. Moreover, these consequences will most strongly affect the poorest countries of the world. According to [4], we have no more than 5-7 years to make effective decisions aimed at combating climate change. During which it is necessary to take radical measures to prevent catastrophic climate changes that can have a negative impact on the human environment.

According to researchers [5], by 2050 it is necessary to reduce the amount of greenhouse gas emissions into the environment by at least half compared to 1990.

The authors [6,7,8,9] believe that the 20-30 years of the XXI century may be critical for ensuring the continued existence of human civilization. During these decades, serious measures should be taken to stabilize greenhouse gas emissions and adjust the development of human society taking into account climate change.

According to estimates [10,11,12], even a 50% reduction in greenhouse gas emissions will not prevent an increase in the average temperature on Earth by 1 – 1.5 $^0$C by 2040. According to [13,14,15,16] if emissions increase like the last 20 years, then by 2060 the average temperature of the Earth may increase by 3-3.5$^0$C.

According to [18,19,20,21,22,23] this may raise the level of the world ocean by 50-60 meters, deserts may appear in Europe, and a significant part of Africa may become little habitable.

According to [24,26], global greenhouse gas emissions of $N_2O$, $SF_6$, $CO_2$, $CH_4$ increased by about 72% between 1970 and 2005. Moreover, $CO_2$ emissions increased by 80% during this period.

According to [27,28,29,30,31], the largest greenhouse gas emissions are inherent in the energy sector of the economy. Their volume in the period from 1970 to 2005 increased by about 1.5 times, the volume of greenhouse gas emissions from agriculture increased by about 26% during this period.

According to [32], the increase in greenhouse gas emissions into the atmosphere is significantly influenced by the increase in global per capita income by 72% over the past 30 years, and global population growth by 32%.

The authors [33,34,35,36,37,38] note that delay in solving the issue of reducing greenhouse gas emissions may require more money and effort in the future, but this will be difficult to achieve based on the available technological capabilities.

A number of researchers [39,40,41,42,43] note that the maximum allowable increase in the average global temperature on Earth should not exceed 2 $^0$C relative to the current value. Otherwise, sudden catastrophic climate changes may occur.

According to [44,45,46,47], climate change due to global warming can significantly exacerbate existing humanitarian, political and economic problems.

According to [48.49], an increase in the average global temperature on earth accompanying an increase in the population will create problems of shortage of drinking water, as well as water for irrigated agriculture.

The authors [50,51,52,53,54,55,56] note that climate change can cause armed conflicts over access to land, natural and water resources.

According to [57,58,59,60], global warming may cause migration flows under the influence of lack of water resources, high population growth rates and rising water levels in the oceans.

The authors [61,62,63,64] indicate that global warming can have a huge negative impact on the implementation of long-term investment projects related to the development of transport infrastructure and housing construction. This is due to the fact that the service life of buildings and transport communications is tens of years. And therefore, already at the stage of their construction, it is necessary to provide for their functioning in the long term, while climatic conditions may change significantly.

According to [65,66,67,68,69,70], global food consumption by the population is increasing from year to year. The absolute areas occupied by arable land are constantly increasing. Thus, in the period from 2000 to 2020, an additional 500 million hectares of land were put into agricultural circulation.

The authors [71,72] note that agriculture accounts for approximately 10-13% of global greenhouse gas emissions. Moreover , the share of agriculture exceeds half of the global emissions of $N_2O$ and $CH_4$ gases .

According to [73], by the middle of this century, the global demand for food may increase, taking into account population growth, by 65%, which means that emissions of greenhouse gases $N_2O$ and $CH_4$ by agricultural enterprises will tend to increase.

According to estimates [74,75,76] in 2050, in order to feed the world's population, it will be necessary to produce 1 billion tons of grain annually, and 400 million tons of meat.

According to [77], there are the following main sources of greenhouse gas emissions in (%) of global emissions (Figure 1):

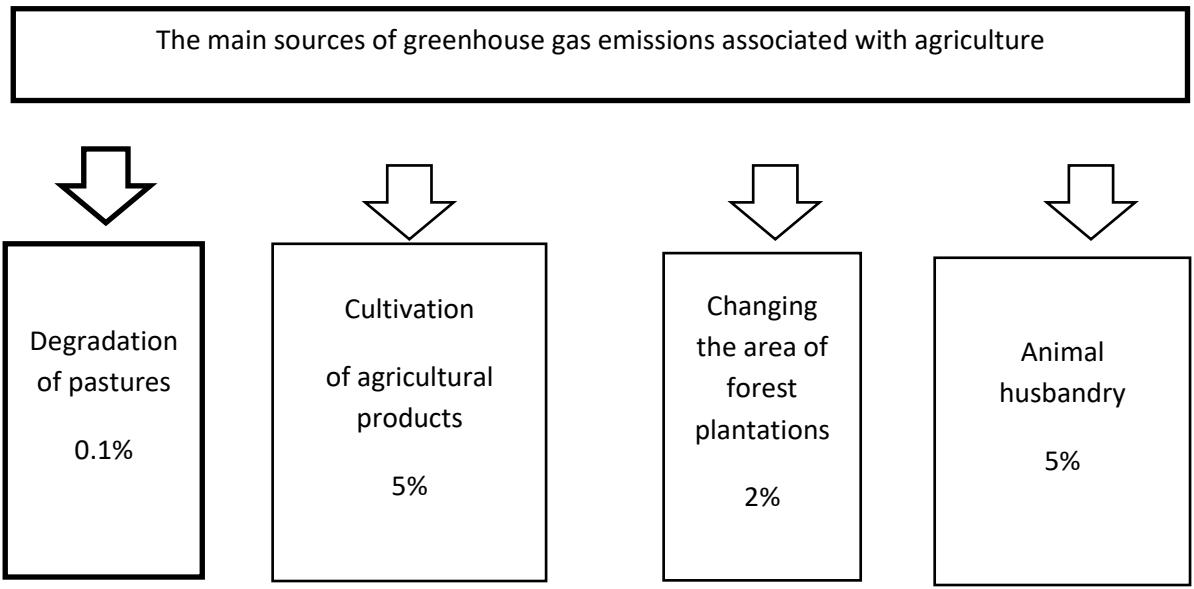

Figure 1. The main sources of greenhouse gas emissions associated with agriculture

According to [78,79,80,81,82] forestry is an effective $CO_2$ sink, it can also be a source of bioenergy for their wood processing waste.

The authors [83,84,85,86,87] note the importance of preserving existing and creating new natural protected areas, which in turn can not only have a beneficial effect on the conservation of various species of flora and fauna, but also make a significant contribution to the fight against the reduction of $CO_2$ concentration in the atmosphere.

A number of researchers note [88.89] that the reduction of $CO_2$ emissions into the atmosphere can be achieved by using agrotechnical technologies that promote the absorption of $CO_2$ by agricultural plants during photosynthesis.

According to [90], this is facilitated by the proper use of fertilizers, the cultivation of improved varieties and hybrids of agricultural crops. Thanks to photosynthesis, carbon enters the stems and leaves of plants and eventually ends up in the soil.

The authors [91] note that the use of the following agrotechnical technologies contributes to an increase in carbon uptake into the soil (Figure 2):

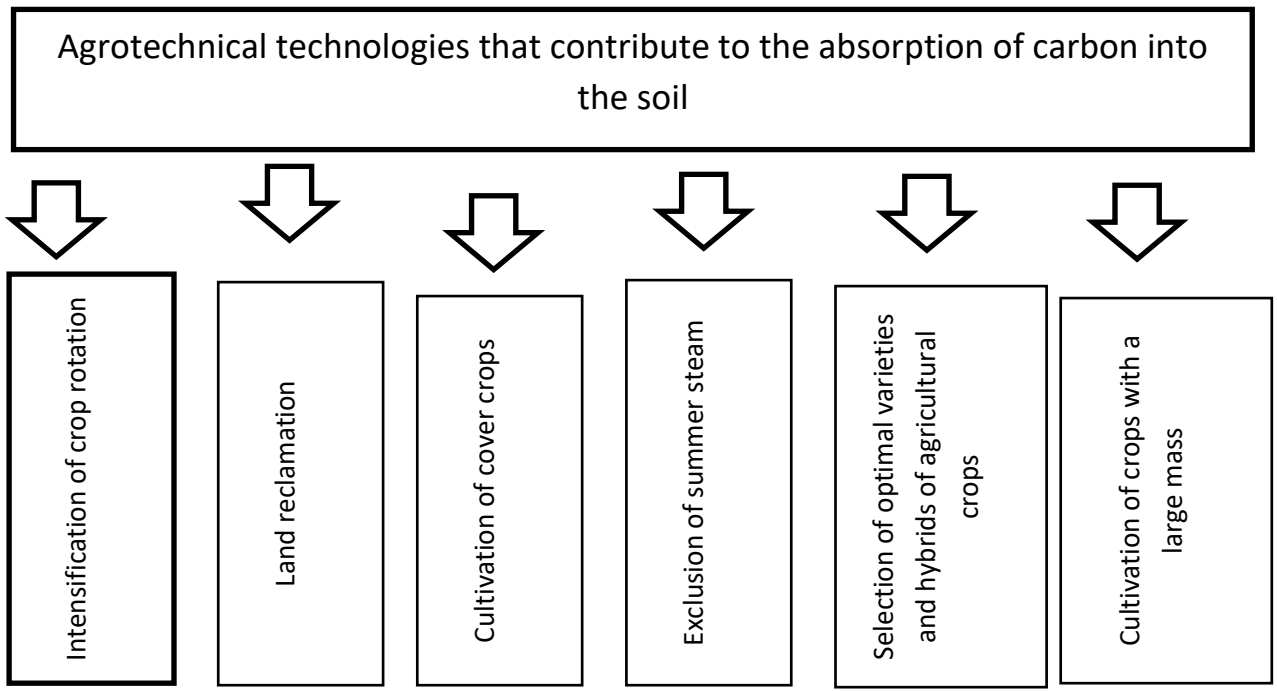

Figure 2. Agrotechnical technologies that contribute to the absorption of carbon into the soil

According to [92], in order to develop and implement an optimal strategy to combat global warming, it is necessary to reach a compromise between the use of land for growing food for an increasing world population, the use of wood production activities and the cultivation of forest plantations to reduce the amount of carbon in the atmosphere.

According to [93], the achievement of food security and adaptation to climate change is possible only as a result of the implementation of an integrated approach to the management of food and agricultural systems. Agriculture has an important role to play as a global carbon sink.

In our opinion, climate-optimized agriculture will solve the problem of increasing agricultural productivity and reducing greenhouse gas emissions. This requires the integration of the agricultural sector of the economy in economic, environmental and social aspects. Which should provide for insurance of agricultural risks, the fight against diseases and pests of agricultural plants, improving environmental education of the population, the introduction of digital information technologies in agricultural production.

Sustainable development of agriculture should be based on the synergetic effect of the use of financial, technical and managerial mechanisms that take into account local conditions.

The development of agrotechnical technologies that increase the carbon content in the soil, in our opinion, will not only increase the level of food security, but also make a significant contribution to the fight against climate change.

According to the author, a significant role in the fight against climate change should be played by regional authorities through the influence of various sectors of the economy Figure 3.

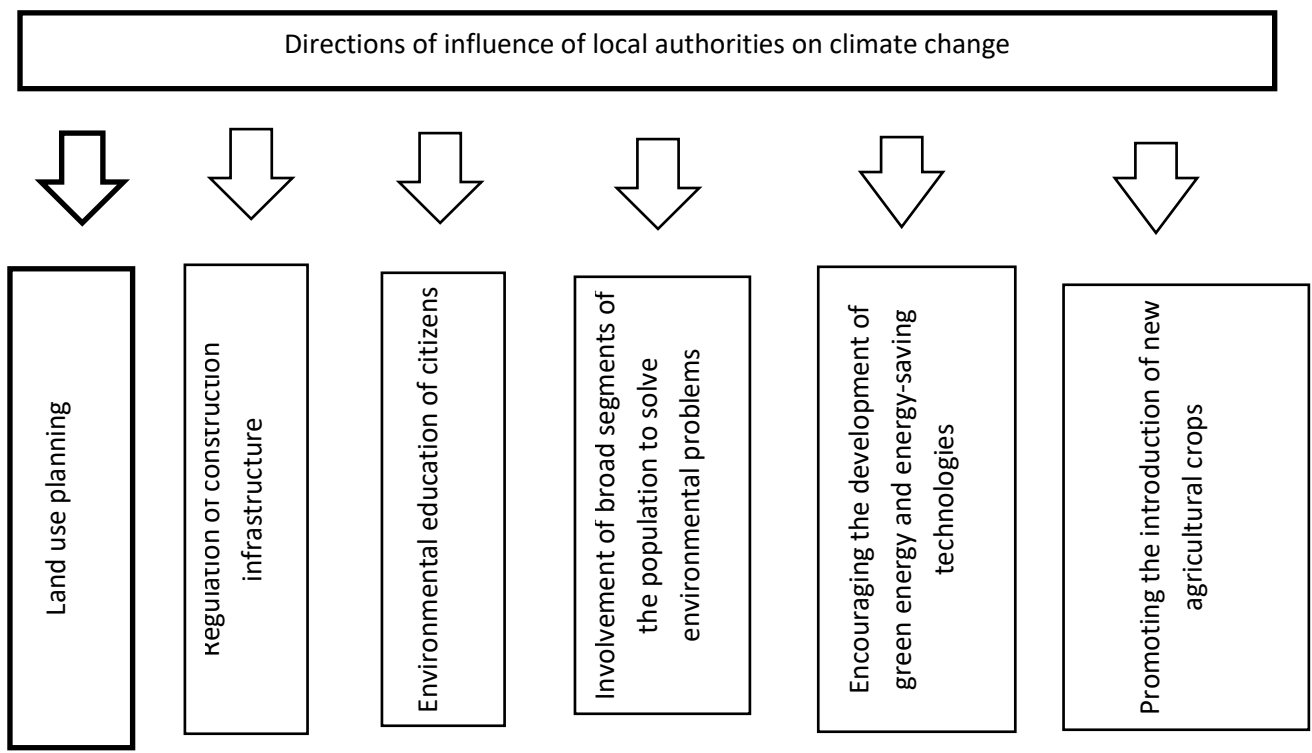

Figure 3. Directions of influence of local authorities on climate change

Trading quotas allows you to significantly revive investment in the environmental industry.

When implementing environmental measures, it is often necessary to sacrifice some economic benefits.

Despite the fact that no one doubts the upcoming climate changes, it is currently necessary to make decisions on how to respond to these changes, which means that investment decisions are made in conditions of uncertainty.

The construction of economic and mathematical models of climate change makes it possible to solve the problem of insufficient data on climate uncertainty to a certain extent. These economic models can be useful in developing plans for the socio-economic development of territories. Activities to combat climate change should take into account the interests of all parties involved in it.

## Methods

Orthogonal polynomials on the orthogonality interval $(a, b) \to R$ are [18] an infinite sequence of real polynomials $p_0(x), p_1(x), p_2(x), \ldots, p_n(x), \ldots$ where each of the polynomials $p_n(x)$ has degree $n$, which are mutually orthogonal in the sense of the scalar product defined in the space $L^2$..

The system of orthogonal polynomials $p_n(x)$, is complete. This means that any polynomial $F(x)$ of degree $n$ on the interval of orthogonality $(a, b) \to R$ can be written as a series:

$$F(x) = \sum_{i=1}^{n} c_i p_i(x) \qquad (1)$$

where $c_i$ - are the expansion coefficients

Let some function $f(x)$ be a continuous function on the segment $[a, b]$, and $\{\omega_k(x)\}$ is a system of continuous orthogonal functions, each of which does not vanish on the entire segment $[a, b]$. It is said that the function $f(x)$ can be decomposed on the segment $[a, b]$ by the orthogonal system of functions $\{\omega_k(x)\}$, if there exists a sequence of numbers $\{a_k\}$ such that the series $\sum_{k=1}^{\infty} a_k \omega_k(x)$ converges and equality (2) takes place:

$$f(x) = \sum_{k=1}^{\infty} a_k \omega_k(x), x \in [a, b] \qquad (2)$$

If the series (1) on the segment $[a, b]$ converges uniformly, then the following expression is valid for the coefficients $a_k$

$$a_k = \frac{\int_a^b f(x) \omega_k(x) dx}{\int_a^b \omega_k^2(x) dx} \quad . \quad k \in N \qquad (3)$$

If the function $f(x)$ is given as a table, then it can be interpolated using a system of continuous orthogonal functions,

$$f(x_i) = \sum_{k=0}^{\infty} b_k \omega_k(x_i), x \in [a, b] \qquad 0 \leq k \leq n$$

$$b_0 \omega_0(x_j) + b_1 \omega_1(x_j) + b_2 \omega_2(x_j) + \cdots + b_n \omega_n(x_j) = f(x_j),$$

where $j = 0, 1, \ldots, n$

Equation (3) can be represented in matrix form:

$$\begin{pmatrix} \omega_0(x_o) & \omega_1(x_o) & \ldots & \omega_n(x_o) \\ \omega_0(x_1) & \omega_1(x_1) & \ldots & \omega_n(x_1) \\ \ldots & \ldots & \ldots & \ldots \\ \omega_0(x_n) & \omega_1(x_n) & \ldots & \omega_n(x_n) \end{pmatrix} \begin{pmatrix} b_0 \\ b_1 \\ \ldots \\ b_n \end{pmatrix} = \begin{pmatrix} f(x_0) \\ f(x_1) \\ \ldots \\ f(x_n) \end{pmatrix} \qquad (4)$$

Where (4) finding $b_o, b_1, ..., b_n$ reduces to solving a system of n-linear equations.

The scalar product of vectors on an n-dimensional Euclidean space $a = (a_1, a_2, ..., a_n)$ and $b = (b_1, b_2, ..., b_n)$ is understood as [100]

$$\langle a, b \rangle = a_1 b_1 + a_2 b_2 + \cdots + a_n b_n \tag{8}$$

Expression (8) is equivalent to

$$\langle\langle a, b \rangle = |a||b|\cos(\theta) \tag{9}$$

Generalize expression (9) for vectors of the space formed by orthogonal functions $\{\omega_k(x)\}$

Suppose there are functions $g(x)$ and $h(x)$

Which can be decomposed into orthogonal polynomials $\{\omega_k(x)\}$

$$g(x_i) = \sum_{k=0}^{\infty} b_k \omega_k(x_i), x \in [a, b] \qquad 0 \le k \le n \tag{10}$$

$$h(x_i) = \sum_{k=0}^{\infty} c_k \omega_k(x_i), x \in [a, b] \qquad 0 \le k \le n$$

Then, by analogy with (9), we can introduce the concept of a scalar product between vectors $g(x_i)$ and $h(x_i)$

$$\langle g(x_i) h(x_i) \rangle = |g(x_i)||h(x_i)|\cos(\theta) \tag{11}$$

$$\cos(\theta)_{ij} = \frac{\sum_{k=1}^{n} b_k^i c_k^j}{\sqrt{[\sum_{k=1}^{n}(b_i^2)]}\sqrt{[\sum_{k=1}^{n}(c_j^2)]}} \tag{12}$$

where $\cos(\theta)_{ij}$ is the cosine of the angle between vectors in the space formed by orthogonal functions $\{\omega_k(x)\}$

In the future we will call the cosine of the angle between vectors in the space formed by orthogonal functions $\{\omega_k(x)\}$ the coefficient of τ-covariance, and we introduce the notation

$$\tau_{ij} = \cos(\theta)_{ij}$$

Obviously, $\tau_{ij} \in [-1, 1]$.

Covariance plays an important role in mathematical statistics

$$cov_{XY} = E[(X - E(X))(Y - E(Y))] = E(XY) - E(X)E(Y) \tag{13}$$

where E – mathematical expectation.

Covariance is one of the characteristics of the distribution of various random variables. Since $cov_{XY} \in [-1,1]$ and the expression $\tau_{ij} \in [-1,1]$. Then the coefficient of τ-covariance, according to the author, can be similarly used to analyze the presence of a statistical relationship between the functions $g(x)$ and $h(x)$.

Orthogonal functions include Hermite polynomials, they are usually expressed as

$$H_n(x) = (-1)^n e^{x^2} \frac{d^n}{dx^n} e^{-x^2} \qquad (13)$$

An important consequence of the presence of orthogonality in Hermite polynomials is the possibility of decomposing various functions over them.

Let's say the function $f(x)$ can be decomposed into a Maclaurin series

$$f(x) = \sum_{n=0}^{\infty} c_n x^n \qquad (14)$$

then, according to [100], the decomposition of the same function by Hermite polynomials looks like

$$f(x) = \sum_{n=0}^{\infty} A_n H_n(x) \qquad (16)$$

, where

$$A_n = \frac{1}{n!} \sum_{k=0}^{\infty} \frac{1}{2^k} \frac{(n+2k)!}{k!} c_{n+2k} \qquad (17)$$

Thus, if $f(x)$ can be decomposed into a Maclaurin series, then it can be decomposed into Hermite polynomials.

## Results

An important problem on the way to reducing greenhouse gas emissions is insufficient research and understanding of climate risks, as well as the development and implementation of a strategy for the development of human society in the near future.

In our opinion, it should be taken into account, first of all, that the implementation of infrastructure projects aimed at climate change should be implemented before the onset of climate change. At the same time, distribute investments between different sectors of the economy for the long term.

In order to successfully implement measures to mitigate climate change and implement a sustainable development strategy, it is necessary first of all to analyze the macroeconomic indicators of the industrial impact of various sectors of the economy on the overall picture of greenhouse gas emissions.

The result of this analysis should be the development of a comprehensive climate policy, the elements of which would be implemented at the local and state levels.

Within the framework of this study, the data were analyzed [94] Table 1.

Correlation analysis is widely used to analyze the interdependence of macroeconomic indicators, its occurrence is largely due to the work [95].

The main disadvantage of correlation analysis according to [96,97,98,99] is the need to subordinate the studied time series to the normal law of probability distribution, we will show this on a concrete example:

Take a sequence of numbers 0;1;2;3;4;5;6;7;8;9;10 , and also a numerical series consisting of the exponents of these numbers exp(0)=1 ; exp(1)= 2.718282 ; exp(2)= 7.389056 ; exp(3)= 20.08554 ; exp(4)= 54.59815 ; exp(5)= 148.4132; exp(6)= 403.4288 ; exp(7)= 1096.633 ; exp(8)= 2980.958; exp(9)= 8103.084; exp(10)= 22026.466

Although the numerical sequences {0,1,2,3,4,5,6,7,8,9,10} and {1, 2.718282, 7.389056, 20.08554, 54.59815, 148.4132, 403.4288, 1096.633, 2980.958, 22026.466} they have a rigid functional relationship with each other, nevertheless, the covariance of these numerical sequences is 0.71687.

In table 2, the Pearson coefficients between the numerical series were presented Forest area (sq. km); CO2 emissions (metric tons per capital); Electric power consumption (kWh per capital);GDP (current US$); Agriculture, forestry, and fishing, value added (% of GDP); Industry (including construction), value added (% of GDP); Exports of goods and services (% of GDP); Imports of goods and services (% of GDP) for the Czech Republic for the period 2000 – 2015

Numerical series of macroeconomic indicators of Forest area (sq. km) ; $CO_2$ emissions (metric tons per capital) ;Electric power consumption (kWh per capital);GDP (current US$); Agriculture, forestry, and fishing, value added (% of GDP); Industry (including construction), value added (% of GDP); Exports of goods and services (% of GDP); Imports of goods and services (% of GDP) for the Czech Republic for the period 2000 – 2015 were decomposed by Hermite polynomials in the interval [0 , 1], where the year 2000 corresponds to the value 0 , and the year 2015 corresponds to the value 1.

GDP (млрд. USD)= -257467542,7*$H_0(x)$+ 575413342,2*$H_1(x)$-458622107,3*$H_2(x)$+

+ 330467806,3*$H_3(x)$-77398076,61*$H_4(x)$ +40935103,59*$H_5(x)$+2798056,589*$H_6(x)$+

+1867609,009*$H_7(x)$+759418,6949*$H_8(x)$+91674,34629*$H_9(x)$+29020,11318*$H_{10}(x)$+

+4933,890807*$H_{11}(x)$+320,7992577*$H_{12}(x)$+114,2072781*$H_{13}(x)$+0,399454495*$H_{14}(x)$+

+0,8192630*$H_{15}(x)$

CO2 emission (metric ton per capita) = 39873530,21*$H_0(x)$-13026678,67*$H_1(x)$+57078760,72*$H_2(x)$+

+37144725,81*$H_3(x)$+11039085,96*$H_4(x)$ +13551517,93*$H_5(x)$+2436261,78*$H_6(x)$ +

+732990,3244*$H_7(x)$+435554,0212*$H_8(x)$-51069,22151*$H_9(x)$+29462,00319*$H_{10}(x)$-

-4502,029741*$H_{11}(x)$+750,9001231*$H_{12}(x)$-98,60386495*$H_{13}(x)$+6,139244274*$H_{14}(x)$+

+0,639471145*$H_{15}(x)$

Agriculture, forestry, and fishing = GDP * [273710246,3*$H_0(x)$-

619677926,4*$H_1(x)$+454615947,5*$H_2(x)$-362411256,6*$H_3(x)$+68672792,73*$H_4(x)$-

+57428724,5*$H_5(x)$-1905829,452*$H_6(x)$ – 5754300,662*$H_7(x)$- 314588,4686*$H_8(x)$-

453484,2664*$H_9(x)$+4508,331813*$H_{10}(x)$- 19286,1096*$H_{11}(x)$+544,8031475*$H_{12}(x)$-

355,6132925*$H_{13}(x)$+6,640203074*$H_{14}(x)$-2,21323152*$H_{15}(x)$]

Industry (including construction) = GDP * [1605793321*$H_0(x)$-3000692173*$H_1(x)$ +

+2873303333*$H_2(x)$-1287329304*$H_3(x)$+482301642,6*$H_4(x)$ +14335223,43*$H_5(x)$ –

-21303521,77*$H_6(x)$ +26264323,72*$H_7(x)$-5493143,094*$H_8(x)$ +1928950,922*$H_9(x)$ –

-229904,529*$H_{10}(x)$+49959,66988*$H_{11}(x)$-3222,494027*$H_{12}(x)$+502,7329109*$H_{13}(x)$-

-12,53372334*$H_{14}(x)$ +1,605166278*$H_{15}(x)$]

Exports of goods and services = GDP * [710681348,6*$H_0(x)$-478952899,3*$H_1(x)$ +

+1229210418*$H_2(x)$ +429359395,8*$H_3(x)$+267948054,4*$H_4(x)$ +269594341,1*$H_5(x)$ +

+25616537,57*$H_6(x)$ +36111846,37*$H_7(x)$ +2683508,507*$H_8(x)$ +1802502,373*$H_9(x)$ +

+174656,188*$H_{10}(x)$+38577,84296*$H_{11}(x)$ +4559,184981*$H_{12}(x)$+366,5995184*$H_{13}(x)$ +

+37,77925541* $H_{14}(x)$+1,458763605*$H_{15}(x)$]

Imports of goods ans service = GDP * [-1159776910*$H_0(x)$+3331728115*$H_1(x)$ –

-1986759199*$H_2(x)$ +2280715751*$H_3(x)$-205085206,7*$H_4(x)$ +377380498,2*$H_5(x)$ +

+63356761,32* $H_6(x)$ + 24589504,25*$H_7(x)$+9541993,567*$H_8(x)$ +923286,1957*$H_9(x)$ +

+433740,5246*$H_{10}(x)$+27166,77358*$H_{11}(x)$+7717,999716*$H_{12}(x)$+506,9179536*$H_{13}(x)$ +

+46,17917842* $H_{14}(x)$+3,632792684*$H_{15}(x)$]

where $H_i(x)$- Hermite polynomials (13)

For vectors of Hilbert space (2) describing numerical series of macroeconomic indicators of Forest area (sq. km) ; $CO_2$ emissions (metric tons per capital) ;Electric power consumption (kWh per capital);GDP (current US$); Agriculture, forestry, and fishing, value added (% of GDP); Industry (including construction), value added (% of GDP); Exports of goods and services (% of GDP); Imports of goods and services (% of GDP) for the Czech Republic for the period 2000 – 2015, the coefficients of τ -covariance (12) were calculated. The final result is presented in Table 3.

Table 1. Numerical series characterizing some macroeconomic indicators of the Czech Republic for the period 2000 – 2015

| Series Name | 2000 | 2001 | 2002 | 2003 | 2004 | 2005 | 2006 | 2007 |
|---|---|---|---|---|---|---|---|---|
| Forest area (sq. km) | 26372,9 | 26392,99 | 26413,08 | 26433,17 | 26453,26 | 26473,35 | 26493,44 | 26513,53 |
| CO2 emissions (metric tons per capita) | 12,01065269 | 12,01181802 | 11,6241027 | 12,04336087 | 12,10540157 | 11,750804 | 11,77860313 | 12,00330783 |
| Electric power consumption (kWh per capita) | 5703,816739 | 5892,172595 | 5894,233119 | 6074,849142 | 6230,398228 | 6357,421095 | 6528,53015 | 6518,217413 |
| GDP (current US$) | 61828166496 | 67808032980 | 82196001051 | 1,0009E+11 | 1,19814E+11 | 1,37143E+11 | 1,56264E+11 | 1,90184E+11 |
| Agriculture, forestry, and fishing, value added (% of GDP) | 3,250528331 | 3,203527086 | 2,659586723 | 2,440027314 | 2,415394613 | 2,284635292 | 2,164955432 | 2,088905575 |
| Industry (including construction), value added (% of GDP) | 33,49120748 | 33,90920025 | 33,2092894 | 32,34225338 | 33,57974959 | 33,61719819 | 34,33321599 | 34,18273662 |
| Exports of goods and services (% of GDP) | 48,09098982 | 48,85651186 | 45,0041286 | 46,72804779 | 57,05783989 | 61,81322686 | 64,87542344 | 66,10076919 |
| Imports of goods and services (% of GDP) | 49,95534908 | 50,1559443 | 46,33052172 | 48,24548106 | 56,43112009 | 59,48473354 | 62,15295843 | 63,67824812 |

| Series Name | 2008 | 2009 | 2010 | 2011 | 2012 | 2013 | 2014 | 2015 |
|---|---|---|---|---|---|---|---|---|
| Forest area (sq. km) | 26533,62 | 26553,71 | 26573,8 | 26595,82 | 26617,84 | 26639,86 | 26661,88 | 26683,9 |
| CO2 emissions (metric tons per capita) | 11,39282817 | 10,64445418 | 10,71659406 | 10,4010178 | 10,09153928 | 9,620257259 | 9,264302843 | 9,400668002 |
| Electric power consumption (kWh per capita) | 6489,126257 | 6139,35206 | 6348,424398 | 6298,727678 | 6304,571923 | 6284,790806 | 6258,891037 | 6268,891066 |
| GDP (current US$) | 2,36816E+11 | 2,07434E+11 | 2,0907E+11 | 2,29563E+11 | 2,08858E+11 | 2,11686E+11 | 2,09359E+11 | 1,88033E+11 |
| Agriculture, forestry, and fishing, value added (% of GDP) | 1,919927972 | 1,756787513 | 1,540596113 | 1,982510992 | 2,251185645 | 2,364771166 | 2,413452542 | 2,211213873 |
| Industry (including construction), value added (% of GDP) | 33,65763346 | 32,98769447 | 33,17075187 | 33,65057382 | 32,9223275 | 32,64607051 | 33,83935076 | 33,7818228 |
| Exports of goods and services (% of GDP) | 62,95164809 | 58,34542981 | 65,54300541 | 70,82186719 | 75,64618657 | 76,05838162 | 81,95427457 | 80,55877811 |
| Imports of goods and services (% of GDP) | 60,79085103 | 54,45181472 | 62,48590613 | 67,04095169 | 70,88267001 | 70,36403543 | 75,6207076 | 74,61688537 |

Table 2. Pearson coefficients between some numerical series characterizing the macroeconomic indicators of the Czech Republic for the period 2000 – 2015

| PIRSON | Forest area (sq. km) | CO2 emissions (metric tons per capita) | Electric power consumption (kWh per capita) | GDP (current US$) | Agriculture, forestry, and fishing, value added (% of GDP) | Industry (including construction), value added (% of GDP) | Exports of goods and services (% of GDP) | Imports of goods and services (% of GDP) |
|---|---|---|---|---|---|---|---|---|
| Forest area (sq. km) | 1 | 0,804915995 | -0,370651112 | -0,170300938 | 0,706956118 | 0,66083114 | 0,551427119 | 0,007525423 |
| CO2 emissions (metric tons per capita) | 0,804915995 | 1 | 0,15519434 | 0,321520301 | 0,902132858 | 0,897826088 | 0,878669481 | 0,501046148 |
| Electric power consumption (kWh per capita) | -0,370651112 | 0,15519434 | 1 | 0,452129202 | 0,394029786 | 0,452129202 | 0,566067763 | 0,925768597 |
| GDP (current US$) | -0,170300938 | 0,321520301 | 0,973262207 | 1 | 0,5736343 | 0,622749449 | 0,713523219 | 0,97723901 |
| Agriculture, forestry, and fishing, value added (% of GDP) | 0,706956118 | 0,902132858 | 0,394029786 | 0,5736343 | 1 | 0,99740355 | 0,972713669 | 0,710828029 |
| Industry (including construction), value added (% of GDP) | 0,66083114 | 0,897826088 | 0,452129202 | 0,622749449 | 0,99740355 | 1 | 0,156492808 | 0,755290483 |
| Exports of goods and services (% of GDP) | 0,551427119 | 0,878669481 | 0,566067763 | 0,713523219 | 0,972713669 | 0,156492808 | 1 | 0,835274197 |
| Imports of goods and services (% of GDP) | 0,007525423 | 0,501046148 | 0,925768597 | 0,97723901 | 0,710828029 | 0,755290483 | 0,835274197 | 1 |

Table 3 Coefficients of τ-covariance between some numerical series characterizing the macroeconomic indicators of the Czech Republic for the period 2000 – 2015

| $\tau_{ij}$ | Forest area (sq. km) | CO2 emissions (metric tons per capita) | Electric power consumption (kWh per capita) | GDP (current US$) | Agriculture, forestry, and fishing, value added (% of GDP) | Industry (including construction), value added (% of GDP) | Exports of goods and services (% of GDP) | Imports of goods and services (% of GDP) |
|---|---|---|---|---|---|---|---|---|
| Forest area (sq. km) | 1 | 0,553868219 | -0,998710332 | -0,727421078 | 0,988376967 | 0,998100945 | 0,729853183 | 0,960387536 |
| CO2 emissions (metric tons per capita) | 0,553868219 | 1 | -0,512904347 | -0,476076253 | 0,427860804 | 0,588709296 | 0,968931039 | 0,301603765 |
| Electric power consumption (kWh per capita) | -0,998710332 | -0,512904347 | 1 | 0,725685841 | -0,994797559 | -0,994873572 | -0,695541451 | -0,973036711 |
| GDP (current US$) | -0,727421078 | -0,476076253 | 0,725685841 | 1 | -0,722099758 | -0,750088327 | -0,610248872 | -0,681398774 |
| Agriculture, forestry, and fishing, value added (% of GDP) | 0,988376967 | 0,427860804 | -0,994797559 | -0,722099758 | 1 | 0,981253718 | 0,623016088 | 0,990518216 |
| Industry (including construction), value added (% of GDP) | 0,998100945 | 0,588709296 | -0,994873572 | -0,750088327 | 0,981253718 | 1 | 0,760629553 | 0,946776578 |
| Exports of goods and services (% of GDP) | 0,729853183 | 0,968931039 | -0,695541451 | -0,610248872 | 0,623016088 | 0,760629553 | 1 | 0,511734557 |
| Imports of goods and services (% of GDP) | 0,960387536 | 0,301603765 | -0,973036711 | -0,681398774 | 0,990518216 | 0,946776578 | 0,511734557 | 1 |

## Discussion

The calculations performed showed that the Pearson coefficient (kPirs) between $CO_2$ emissions (metric tons per capital) and Electric power consumption (kWh per capital) is kPirs =0.15519434, and the coefficient of τ -covariance between $CO_2$ emissions (metric tons per capital) and Electric power consumption (kWh per capital) is τ=0.512904347. The Pearson correlation coefficient between $CO_2$ emissions (metric tons per capital) and GDP (current US$) is kPirs =0.321520301, the coefficient of τ -covariance between $CO_2$ emissions (metric tons per capital) and GDP (current US$) is τ=0.476076253. The Pearson correlation coefficient between $CO_2$ emissions (metric tons per capital) and Agriculture, forestry, and fishing, value added (% of GDP) is kPirs =0.902132858, the coefficient of τ-covariance between $CO_2$ emissions (metric tons per capital) and Agriculture, forestry, and fishing, value added (% of GDP) is τ=0.427860804. Pearson correlation coefficient between $CO_2$ emissions (metric tons per capital) and Industry (including construction), value added (% of GDP) is kPirs=0.902132858, coefficient of τ -covariance between $CO_2$ emissions (metric tons per capital) Industry (including construction) is τ=0.588709296. The Pearson correlation coefficient between $CO_2$ emissions (metric tons per capital) and GDP Imports of goods and services (% of GDP) is kPirs =0.501046148, the coefficient of τ - covariance between $CO_2$ emissions (metric tons per capital) and Imports of goods and services (% of GDP) is τ=0.301603765

According to the Pearson coefficients calculated by us, there is a strong correlation between $CO_2$ emissions and exports of goods and services Pears coefficients kPirs =0.888669481, $CO_2$ emissions and the volume of engineering production kPirs =0.897826088, $CO_2$ emissions and agricultural production kPirs= 0.888669481, as well as $CO_2$ emissions and GDP, kPirs= 0.973262207.

According to the coefficients of the τ -covariance calculated by us, there is no strong correlation with the import of goods and services τ = 0.301603765, the volume of machine-building production τ = 0.588709296 and agricultural production τ = 0.427860804, as well as the size of GDP τ = 0.476076253.

In [94] it is indicated that in the period from 2000 to 2015, the Czech Republic's GDP increased from $61.83 billion to $ 188.03 billion, agricultural production increased $2.01 billion to $4.16 billion, mechanical engineering production increased $ 20.71 billion to $63.52 billion, exports of goods and services increased by $29.73 billion to $151.48 billion, imports of goods and services increased by $30.89 billion to $140.30 billion,

At that time, the amount of $CO_2$ emissions in the period from 2000 to 2015 decreased from 12 metric tons per capital to 9.4 metric tons per capital. Since they move in different directions, the Pearson coefficient should not be close to one.

Considering that the indicators of GDP, agricultural production, machine-building production, exports of goods and services, imports of goods and services grew in the period from 2000 to 2015, and the amount of $CO_2$ emissions decreased on the contrary, in our opinion, it is the coefficients of $\tau$ -covariance that describe the interdependence of these macroeconomic indicators more precisely, since their value is somewhat below the Pearson coefficients.

## Conclusions

The maximum effect of the strategy to combat climate change can be achieved if we focus on meeting the interests of society. The actions of the authorities will be more effective if they have the support of the population. Various legislative changes will be more effective if they take into account the social problems that exist in society.

Public support for measures to combat climate change is possible provided that the climate problem is understood by broad segments of the population.

Many projects aimed at combating climate change are not related to each other, so it is very important to properly allocate the available financial resources between them.

The use of a mathematical apparatus based on orthogonal polynomials and the coefficients of $\tau$ -covariance calculated with their help allows us to investigate the interdependence of macroeconomic indicators that do not have a normal distribution, for which correlation analysis based on Pearson coefficients does not give the necessary accuracy of the result.

Orthogonal Hermite polynomials can be used as such polynomials.